\documentclass{article}

\usepackage{arxiv}

\usepackage[utf8]{inputenc} % allow utf-8 input
\usepackage[T1]{fontenc}    % use 8-bit T1 fonts
\usepackage{hyperref}       % hyperlinks
\usepackage{url}            % simple URL typesetting
\usepackage{booktabs}       % professional-quality tables
\usepackage{amsfonts}       % blackboard math symbols
\usepackage{nicefrac}       % compact symbols for 1/2, etc.
\usepackage{microtype}      % microtypography
\usepackage{lipsum}
\usepackage{graphicx}
\graphicspath{ {./images/} }
\usepackage{amsmath}
\usepackage{subcaption}

\title{Language Model Teams \\
as Distributed Systems}

\author{
  Elizabeth Mieczkowski$^{1}$\thanks{Code available at \url{https://github.com/emieczkowski/distributed-llm-teams}} \And
  Katherine M. Collins$^{1,2,3}$ \And
  Ilia Sucholutsky$^{4}$ \And
  Natalia Vélez$^{1}$ \And
  Thomas L. Griffiths$^{1}$
  \\
  \\
  $^{1}$Princeton University \quad
  $^{2}$Massachusetts Institute of Technology \quad
  $^{3}$University of Cambridge\quad
  $^{4}$New York University
}

\begin{document}
\maketitle
\begin{abstract}
Large language models (LLMs) are growing increasingly capable, 
prompting recent interest in LLM \textit{teams}. Yet, despite increased deployment of LLM teams at scale, we lack a principled framework for addressing key questions such as when a team is helpful, how many agents to use, how structure impacts performance --- and whether a team is better than a single agent. Rather than designing and testing these possibilities through trial-and-error, we propose using distributed systems as a principled foundation for creating and evaluating LLM teams. We find that many of the fundamental advantages and challenges studied in distributed computing also arise in LLM teams, highlighting the rich practical insights that can come from the cross-talk of these two fields of study. %providing support for connecting these two fields of study. 

\end{abstract}

% keywords can be removed
\keywords{Large language models \and collaboration \and LLM teams \and distributed computing}

\section{Introduction}

\textit{``Adding manpower to a late software project, makes it later.'' -- Frederick P. Brooks, The Mythical Man Month \cite{brooks1995mythical}}

The quest for artificial intelligence (AI) has long captivated human minds \cite{turing1950}, and the remarkable capabilities of large language models (LLMs) have sparked widespread excitement that we are approaching this goal. Yet, single models remain fundamentally limited in memory, context, and ability. A natural response is to compose them into teams, or collections of interacting agents that can divide work, communicate, and pool their individual resources. The promise is compelling: just as people achieve far more together than alone \cite{henrich2015secret, tomasello2005understanding}, teams of LLM agents could extend capabilities beyond those of any single monolithic model.  

However, achieving this promise requires addressing several questions: 
when do teams help, when do they hurt, and how should they be designed and deployed to maintain performance and efficiency?
Failing to appropriately address these questions can have substantial implications. Individual LLM calls are already resource-intensive, demanding substantial compute and energy to query, and racking up hefty monetary costs to users. These resource demands grow rapidly as agents exchange messages and iterate on shared work. Without careful orchestration to address coordination challenges raised by teams~\cite{bhatt2025should}, agents may overwrite one another, produce redundant outputs, conflict in decisions, propagate errors through chains of reasoning, and reinforce incorrect conclusions through mutually sycophantic exchanges. 

The journey from single- to multi-agent LLM systems mirrors the historical evolution of computing itself \cite{ramon2021survey, van2017distributed}. 
% As such, we can anticipate problems that may arise from transitioning from a single processor to many.
Early computing systems relied on single processors, but as demands for scale grew, engineers turned to distributed architectures that combined many machines to achieve greater capacity and robustness
(Figure~\ref{fig:introductory}A). However, this also brought about new challenges that have defined core research topics in distributed computing for decades, such as issues of coordination, consistency, and task assignment
(Figure~\ref{fig:introductory}B)
. LLM teams appear to be following a similar path; as single models approach limits in context and reliability, there is increasing interest in assembling teams of interacting agents to overcome their individual constraints. However, the design and deployment of these teams have revealed many additional complexities, such as limited scaling depending on the task \cite{yang2026understanding, kim2025towards}, reduced collective problem-solving abilities \cite{pappu2026multi}, and amplifications of errors, false confidence, and confusion \cite{shapira2026agents}.
Viewing LLM teams through the lens of distributed systems (Figure~\ref{fig:introductory}C) both explains observed inefficiencies and generates actionable solutions, providing principled design guidelines, testable hypotheses, and normative benchmarks for evaluating team behavior across varying tasks and conditions. Rather than optimizing LLM teams through trial and error, this framework offers a rigorous foundation for understanding where coordination breaks down and how to improve it.

In this paper, we outline concrete guidelines from distributed computing for designing and evaluating efficient LLM teams, backed by initial empirical demonstrations. We begin by reviewing existing approaches to LLM team design and their limitations (Section 2), then establish the formal correspondence between LLM teams and distributed systems (Section 3). Section 4 provides an empirical test of this analogy. We first show that LLM teams collaborating in simplified settings with pre-assigned tasks exhibit efficiency gains that generally mirror classic scaling laws in distributed computing. We then extend this evaluation to self-coordinating decentralized teams, finding that LLM teams exhibit the same challenges, such as consistency conflicts, architectural tradeoffs, communication overhead, and stragglers, that distributed systems theory anticipates. Finally, Section 5 outlines future directions and limitations of this framework.

These connections position distributed computing as a principled foundation for designing and evaluating LLM teams. As these teams are deployed, failure to address these scalability and coordination concerns risks more than inefficient computation. Poorly designed systems may waste enormous computational resources while producing unreliable outputs. Because these failures emerge from interactions among agents, they can be difficult to diagnose or correct without principled design frameworks. A formal foundation for LLM team architectures -- and deciding whether to implement a team in the first place -- therefore offers a path toward deploying these systems efficiently, robustly, and responsibly at scale.

\section{Background}

\subsection{The rise of LLM teams}

LLM teams, or multi-agent systems in which multiple language model agents coordinate through task decomposition, role specialization, and inter-agent communication, are increasingly being deployed in both research and production. Experimental studies show that sufficiently capable LLM agents can successfully cooperate in simple economic settings like common-pool resource problems \cite{piatti2024cooperate}. Researchers have argued that such teams may be particularly valuable for scientific discovery, where distributing subtasks across agents can increase effective capacity, enable parallel exploration, and reduce individual errors through cross-checking \cite{zhangposition}. Several frontier labs are deploying agentic coding teams that allow users to coordinate multiple LLM instances working together on shared coding tasks \cite{qian2024chatdev, anthropic2025multiagent}.
Despite this enthusiasm, there is a growing need to evaluate these systems rigorously, both in terms of individual models \cite{rahwan2019machine} and their emergent collective behaviors \cite{willis2026evaluating}.
% More broadly, these developments reflect a shift from treating LLMs as isolated tools toward viewing them as interacting components embedded in larger sociotechnical and computational systems \cite{rahwan2019machine, park2023generative}. % NV: I feel like this paragraph is missing a bit of a "turn" at the end, to set up the transition to the next section. e.g., despite intense interest in the promise of LLM teams, there is growing evidence that achieving that promise is not trivial

\subsection{Benefits, limitations, and risks of LLM teams}

There is growing empirical evidence that the performance of multi-agent LLM teams relative to single agents is mixed and highly task-dependent.

On one hand, teams can outperform individual agents by aggregating complementary reasoning paths, enabling cross-checking, and distributing long-horizon context across multiple agents. LLM teams have demonstrated impressive scientific discovery capabilities, including the design of novel nanobodies \cite{swanson2025virtual} and autonomous hypothesis-generation pipelines for chemistry \cite{boiko2023autonomous}.
Controlled studies further show that ensembles of interacting LLM agents can improve accuracy by enabling diversity and debate \cite{bhattacharyyasocial, du2024improving}, and by decomposing contexts that would exceed a single agent's limits across multiple agents \cite{zhang2024chain}.
Early benchmarks reinforced this optimistic view: increasing the number of agents answering a question and asking them to vote upon the correct output often increased accuracy \cite{li2024more}.

However, subsequent work has identified important limitations and failure modes. Performance gains from adding agents are neither linear nor guaranteed; instead, they depend strongly on factors such as agent heterogeneity and coordination structure \cite{pappu2026multi, yang2026understanding}. Multi-agent systems tend to benefit on parallelizable or decomposable tasks but can degrade on sequential or tightly coupled workflows due to communication overhead and coordination errors \cite{kim2025towards}. Moreover, LLM agents are often biased towards being helpful and agreeable rather than honest \cite{liu2024large}, which may suppress productive disagreement and reduce collective problem-solving quality \cite{pappu2026multi}. Interacting agents can also amplify misinformation and adversarial behavior, uncontrollably consuming resources and reporting task completion despite failures  \cite{shapira2026agents}. 

Together, these findings suggest that the advantages of LLM teams are conditional rather than universal, motivating a deeper analysis of when and why teams outperform single agents.

\subsection{Existing design approaches}
\label{exampledesigns_bkgd}
A major line of work seeks to improve LLM teams by drawing inspiration from human collaboration and organizational design. The challenges faced by LLM teams resemble many that we regularly face when collaborating with people (e.g., dividing tasks, coordinating actions, communicating). 
%As such, the design of LLM teams has drawn heavily on principles from human teams and organizations. 
Frameworks such as MetaGPT, AgentVerse, and AutoGen explicitly structure agents into familiar roles (e.g., planner, engineer, reviewer) \cite{hong2023metagpt, chen2023agentverse, wu2024autogen}, and hierarchical coordination patterns resembling managerial workflows can improve multi-agent performance \cite{guo2024embodied, hu2025owl}.
Beyond role-based hierarchies, other work explores alternative coordination structures. Some approaches emphasize deliberative collectives, in which multiple agents debate or critique one another to improve reasoning accuracy \cite{bhattacharyyasocial, du2024improving}, though the gains from debate are sensitive to team design and communication \cite{wu2025can}. Separate lines of work treat the structure of multi-agent communication as its own design variable, finding that topology substantially affects scaling and performance \cite{qian2024scaling}. More recent work aims to automate topology selection by learning task-adaptive communication graphs \cite{zhang2024g}. 

\subsection{The need for a formal framework}

Despite rapid progress in building multi-agent LLM systems, there remains limited systematic understanding of when and why teams outperform individual models, or how their structure should be designed for a given task. Existing design approaches largely propose coordination patterns heuristically, often by analogy to human organizations or empirical benchmarking, rather than deriving them from formal properties of tasks or workflows. Evaluations also focus predominantly on accuracy, leaving questions of efficiency, cost, and robustness largely unaddressed. This gap makes it difficult to anticipate when LLM teams will improve performance versus introduce coordination overhead, redundancy, or failure modes. A formal framework transforms this problem. Rather than discovering failure modes empirically after deployment, such a framework allows us to derive conditions under which teams are expected to succeed or struggle, turning design decisions that are currently made by intuition into ones that can be reasoned about, compared, and tested.

\begin{figure}[t]
    \centering
    \includegraphics[width=\linewidth]{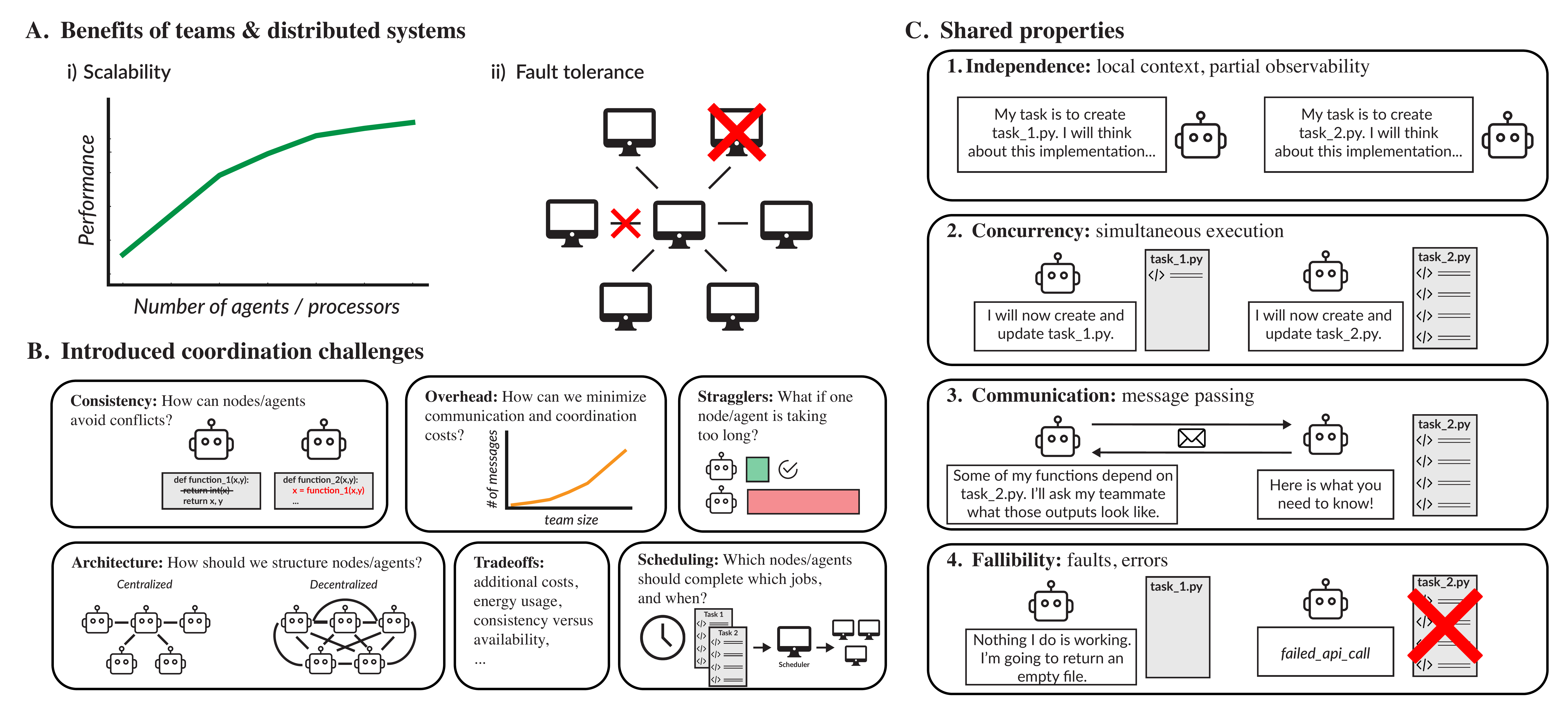}
    \caption{\textbf{LLM Teams as Distributed Systems.} Distributed computing provides a principled framework for analyzing and designing LLM teams. \textbf{A.} Both LLM team and distributed systems research pursue similar goals: leveraging \textit{scalability} to improve performance and achieving \textit{fault tolerance} through mechanisms such as redundancy, replication, and consensus. \textbf{B.} At the same time, LLM teams inherit fundamental complexities familiar from distributed systems but absent in single-agent settings, including consistency conflicts, architectural trade-offs, communication overhead, stragglers, task scheduling, and increased compute, energy, and monetary costs. \textbf{C.} LLM teams share four core properties with distributed systems: \textit{independence} (each agent or node operates on local context without automatic access to global state); \textit{concurrency} (multiple agents or nodes execute tasks simultaneously); \textit{communication} (information is exchanged through message passing); and \textit{fallibility} (agents or nodes may produce errors or undergo faults). }
    \label{fig:introductory}
\end{figure}

\section{A framework for evaluating LLM teams as distributed systems}

Single language model agents have limited capacity: context windows bound how much information they can access at once, memory limits what can be retained, and reasoning, execution, or tool use all require time. Additionally, single agents are prone to faults such as hallucinations, missing relevant context, or failing to respond, all of which disrupt the task they are responsible for. 
As a result, no single agent can do everything at once, and we might not want it to. 

These challenges are not unique to LLM agents. 
Individual machines suffer from similar constraints; they are bounded by memory, time, and unpredictable failures or crashes. These exact problems have driven decades of research in distributed systems, suggesting that its theories offer a natural and rigorous foundation for understanding LLM teams.

\subsection{Shared properties}

To begin, let us make this analogy precise. An LLM team shares four key properties of the core structure of a distributed system (Figure~\ref{fig:introductory}C).

\begin{enumerate}
    \item \textbf{Independence:} LLM agents are independent, maintaining their own local contexts with only partial observability of the state of the broader task and team. Similarly, nodes in a distributed system operate with local memory and have no global clock or state. In both cases, agents or nodes might be \textit{heterogeneous}, having different capabilities due to distinct prompts, base models, or processing times. 
    \item \textbf{Communication:} LLM teams coordinate through communication. Rather than sharing state directly, they exchange messages such as prompts to divide and integrate work. This mirrors how nodes in a distributed system exchange packets to coordinate computation. 
    \item \textbf{Concurrency:} In an LLM team, multiple agents are working on tasks simultaneously. This parallelism can increase speed and coverage, but also introduces major coordination problems: agents may act using stale information, produce conflicting outputs, or overwrite each other's progress. Distributed systems face the same problems when operating in parallel on shared data. Ensuring that all nodes maintain consistency requires synchronization protocols that determine how and when nodes exchange updates and commit results. 
    \item \textbf{Fallibility:} LLM agents can hallucinate, stall, or produce incorrect outputs that propagate through a team. Nodes in a distributed system can crash, fall out of sync, or return corrupted results. In both cases, the system must be designed to tolerate these faults gracefully. 
\end{enumerate}

These four properties establish a structural correspondence between LLM teams and distributed systems, offering a new analytical scaffold. This is not to say that LLM teams conform to every assumption made in distributed computing. For example, communication in LLM teams occurs in natural language rather than fixed, formally specified protocols, making it ambiguous or shaped by pragmatic interpretation. 
Similarly, traditional distributed systems models often assume well-defined failure modes, whereas LLM failures can be semantic and probabilistic.
We believe that these mismatches sharpen the usefulness of this analogy rather than undermine it. 
Where the analogy holds, we inherit decades of existing theories by which we can generate concrete predictions about LLM team behavior. 
Where it breaks down, the gap itself becomes informative, revealing where new theory is needed and providing a baseline against which deviations become visible and measurable. 

\subsection{Predicting LLM team performance using this framework}

Applying this framework to LLMs yields specific predictions about how behavior should change as a result of task and team structure. Here, we test two specific predictions: (1) how task structure affects the efficiency gains achievable through division of labor, and (2) how team architecture shapes coordination costs.

First, this framework yields specific predictions about how performance should change with team size depending on the task. Work in distributed systems has formally characterized how efficiency should scale depending on how much of a task can be executed simultaneously by multiple nodes  \cite{amdahl1967validity, gustafson1988reevaluating, gunther2008general}. We predict that the same efficiency limits should apply to LLMs, such that teams will exhibit stronger efficiency gains when tasks are highly parallel compared to ones with serial dependencies. 

This framework also yields predictions about the tradeoffs between different team architectures. Centralized architectures, in which one node delegates tasks and integrates results, reduce overhead by routing communication through fewer channels. They also improve consistency by enforcing that only one agent updates shared resources or dependencies at a time.
However, because task assignments are fixed, these systems are also vulnerable to variability within nodes; one slow agent (or ``straggler'') can delay the team as a whole.
Decentralized architectures mitigate this risk by allowing tasks to be assigned dynamically, but at the cost of greater coordination overhead and elevated risk of conflicts when multiple agents concurrently modify shared resources. Together, these principles motivate the prediction that centralized and decentralized teams have distinct strengths and vulnerabilities: decentralized teams should exhibit greater coordination overhead (e.g., number of messages, idle rounds) and conflict rates (e.g., file conflicts, rewrites, intermediate test failures), while centralized teams should show greater vulnerability to stragglers.

%Here, we present evidence for the benefits of this conceptual framework by empirically testing two parallels between distributed systems and LLM teams: whether performance in LLM teams scales with group size in ways consistent with scaling laws from distributed computing (Experiment 1), and whether centralized and decentralized LLM teams suffer coordination challenges analogous to those faced by distributed systems (Experiment 2). 
We tested these correspondences by assigning teams of LLM agents to three collaborative coding tasks: implementing a math utilities library, analyzing simulated data, and rendering an SVG file (for more details, see Appendix~\ref{empiricalsetup}). Teams were composed of 1, 2, 3, 4, or 5 homogeneous agents drawn from a single base model (Claude-Sonnet-4-6, Gemini 3-Flash, or GPT-5.2). In each of these tasks, agents were presented with twenty programming subtasks, each specifying a description, required output, and corresponding test file. There were three task structures per domain: a \textbf{highly parallel} task in which eighteen subtasks were mutually independent, a \textbf{mixed} task with ten subtasks forming a sequential dependency chain and remaining ten independent, and a \textbf{highly serial} task with sixteen interdependent subtasks. We tested two assignment schemes. In Experiment 1, coordination challenges were minimized by pre-assigning tasks to different agents to isolate the effects of task structure on scalability. In Experiment 2, this was relaxed by prompting agents to both select which tasks they would perform and then complete them. 

\section{Results}

\subsection{Amdahl's Law predicts scalability in LLM teams}

A central motivation for distributed systems is scalable performance: if large-scale computing tasks are decomposed across many nodes, increasing system size can improve efficiency in terms of completion times or throughput. However, decades of research demonstrate that scalability is often neither linear nor guaranteed \cite{amdahl1967validity, gustafson1988reevaluating, gunther2008general}. Even before accounting for coordination overhead, classical scalability laws demonstrate that performance gains depend primarily on \textbf{parallelizability}, or the extent to which a task can be executed concurrently. 

The ability for multiple nodes to execute a task in parallel is constrained by the inherent structure of the task itself. Bottlenecks due to locking, sequential dependencies between subtasks, shared memory accesses, or resource contention force nodes to wait, limiting parallelism \cite{amdahl1967validity, hill2008amdahl}. Amdahl's Law formalizes how these constraints limit speedup $S$ with $s$ available processors under fixed workloads:

\begin{equation}
    S_{\text{latency}}(s) = \frac{1}{(1-p) + \frac{p}{s}},
\end{equation}
where $p$ is the parallelizable fraction of the workload and $1-p$ is inherently serial. For example, when 95\% of a task is parallelizable, a speedup of 20$\times$ is achievable, and even thousands of nodes can contribute before efficiency plateaus. In contrast, if only 50\% of a task can be parallelized, then even with infinitely many processors and no coordination overhead, the maximum speedup is only 2$\times$, and that plateau is reached with far fewer nodes.

The same efficiency limits should apply to LLM teams. Distributing work across multiple agents can improve performance only to the extent that the underlying task is parallelizable: highly decomposable tasks with independent components should exhibit larger speedups than those with tight interdependencies or contention. Emerging evidence supports this prediction, showing that LLM teams perform better on decomposable than sequential tasks \cite{kim2025towards}. This raises a natural question: can classical scalability laws such as Amdahl’s Law mechanistically explain when scaling LLM teams yields genuine efficiency gains?

\begin{figure}[t]
    \centering
    \includegraphics[width=\linewidth]{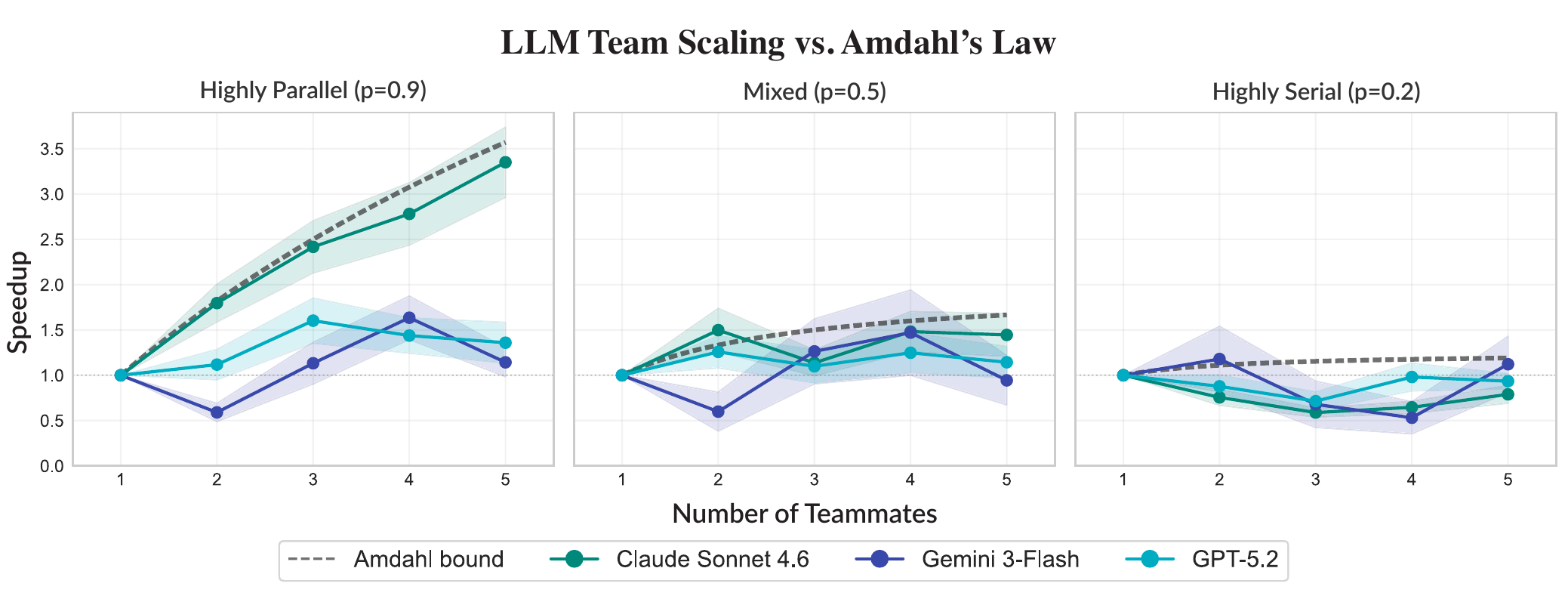}
    \caption{\textbf{Scalability.} A comparison of LLM team scalability to Amdahl's Law, which predicts theoretical speedup based on the proportion of serial dependencies in a task. Teams of agents were given preassigned tasks of three types (coding a math utilities library, creating a data analysis pipeline, and SVG rendering) and three dependency structures (parallel, mixed, or serial). Each trial type was repeated five times to account for variance in API latency, and efficiency was measured using wall-clock time in seconds. Speedup represents how much faster a team completed their task compared to the one-agent baseline. Highly parallel tasks generally benefited more from scaling team size than mixed or serial tasks, as predicted by Amdahl's Law, although results depended on model type. }
    \label{fig:amdahllaw}
\end{figure}

\begin{figure}[t]
    \centering
    \includegraphics[width=\linewidth]{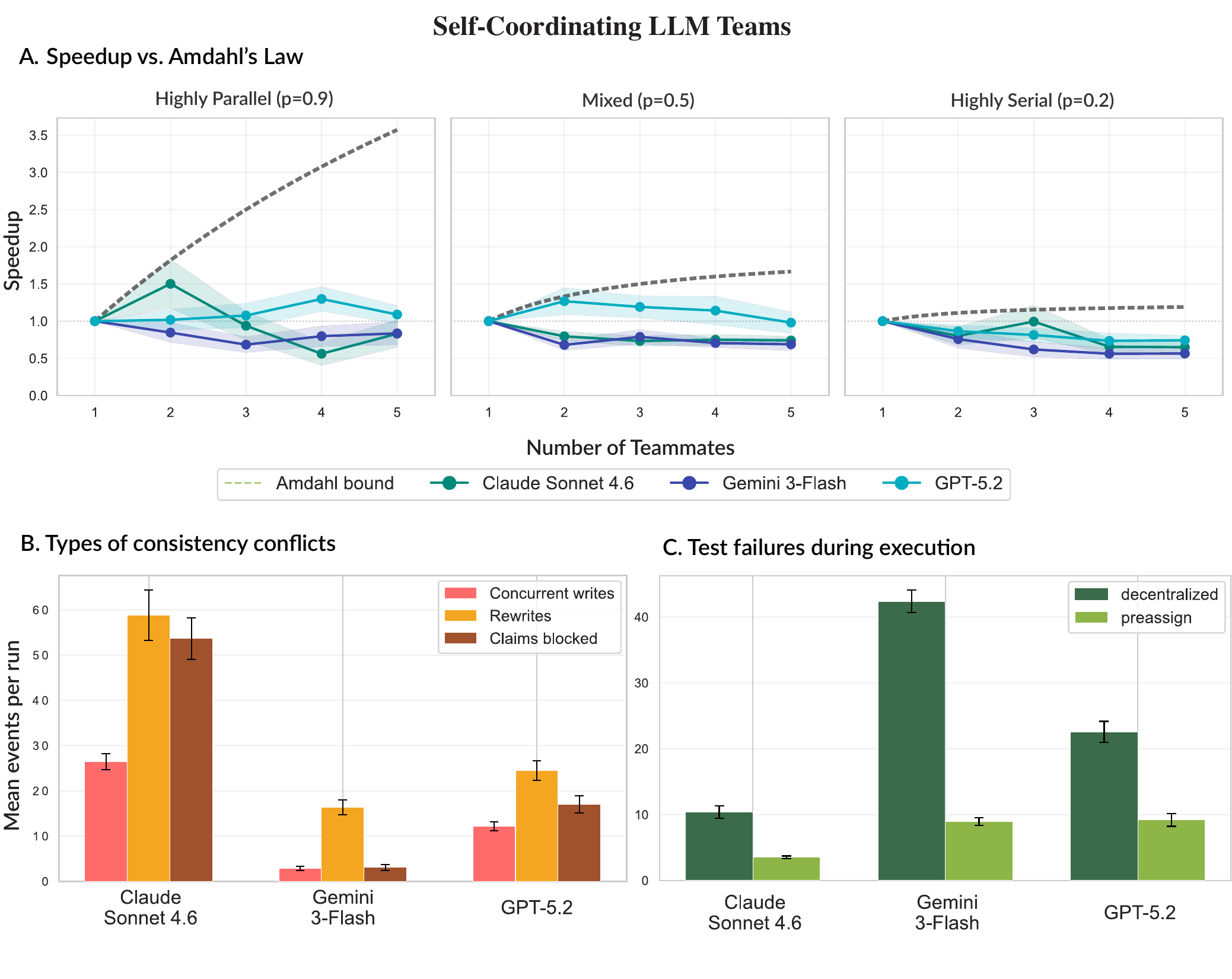}
    \caption{\textbf{Self-coordinating (decentralized) LLM teams.} In Experiment 2, agents needed to not only complete tasks but also decide on assignments themselves. \textbf{A. Scalability:} Speedup is substantially lower for self-coordinating than preassigned teams due to consistency conflicts and communication overhead. This difference is especially stark for highly parallel tasks. \textbf{B. Consistency conflicts:} In self-coordinating teams, agents exhibit conflicts like writing to the same file simultaneously (\textit{pink}), rewriting a file that another agent previously wrote (\textit{yellow}), and attempting to complete a function before its dependencies have been finished (\textit{brown}). These problems do not arise when tasks are preassigned by a central coordinator. \textbf{C. Test failures:} Failed test cases per round reveal that decentralized teams exhibit higher rates of intermediate failure due to these conflicts. }
    \label{fig:amdahllaw_decentralized}
\end{figure}

We tested this question directly in Experiment 1 using our collaborative coding platform. Parallelizability was manipulated by varying prespecified dependencies between subtasks, resulting in a set of highly parallel, mixed, and sequential tasks. To simplify coordination, we preallocated tasks to agents to isolate the effect of task structure on scalability. We measured end-to-end wall-clock completion time as a function of team size, and calculated speedup as the ratio of single-agent to N-agent completion time, $S(N)=T(1)/T(N)$, allowing direct comparison between observed speedups and the limits predicted by Amdahl’s Law.

Figure~\ref{fig:amdahllaw} shows that Amdahl’s Law provides a clear bound on the efficiency gains achievable by LLM teams. As predicted, speedup differed across parallelizability conditions (Kruskal-Wallis: $H=61.4$, $p<0.001$). Highly parallel tasks benefited most from distributing work across agents, where independent subtasks allowed clean partitioning and balanced workloads. Tasks with mixed dependencies generally exhibited less speedup when the number of agents increased, and tasks with highly serial dependencies gained relatively no improvement. Pairwise comparisons confirmed the predicted ordering ($p_{0.9} > p_{0.5}$ MWU $U = 18493$, $p<0.001$; $p_{0.5} > p_{0.2}$ $U = 17847$, $p<0.001$). However, even in the highly parallel condition, speedup remained significantly below the Amdahl bound (Wilcoxon signed-rank, $M=2.19\times$, $p<0.001$). GPT-5.2 and Gemini 3-Flash  were the main drivers of this effect; Claude Sonnet 4.6 alone did not significantly plateau below the bound ($p=0.45$).

\subsection{Tradeoffs arise with architectural choices}

Distributed systems also offer insight into the coordination challenges that arise when tasks and information are divided between nodes. Architectural choices introduce tradeoffs between efficiency, consistency, and robustness.
System architectures range from \textbf{centralized}, where a single coordinator manages shared state and task assignment, to \textbf{decentralized}, where components coordinate through local decisions and communication \cite{van2017distributed}. Each approach has well-known trade-offs. Centralized coordination simplifies consistency challenges, but introduces communication bottlenecks and single points of failure. Decentralized coordination improves robustness and parallelism, but increases negotiation overhead, redundancy, and the risk of conflicting decisions.

In LLM teams, similar decisions must be made about whether a central \textit{leader} (whether human or LLM) should assign tasks to agents, or whether agents will autonomously decide how to allocate work among themselves. %To examine how coordination structure shapes efficiency in practice, we directly compare these two regimes by examining differences between Experiments 1 and 2. Experiment 1 mirrors a centralized structure, while Experiment 2 reflects a decentralized one in which agents must autonomously allocate work.
In Experiment 2, we tested the performance of decentralized LLM teams in our suite of collaborative coding tasks. Rather than pre-assigning tasks to agents, agents were given the ability to self-claim them while communicating with their teammates. We again tested teams of 1, 2, 3, 4, or 5 agents to directly compare performance of these self-coordinating teams to the centralized ones in Experiment 1.

Figure~\ref{fig:amdahllaw_decentralized}A shows that decentralized coordination often reduces efficiency relative to the centralized teams in Figure~\ref{fig:amdahllaw}. Across all runs, preassigned teams achieved significantly higher speedup than decentralized teams (Mann–Whitney $U = 155523$, $p < 0.001$), with median speedups of 
$1.36\times$ versus $0.88\times$ respectively. This pattern held within each model individually (all $p \leq 0.01$). In the following sections, we examine whether the predicted costs of decentralization---specifically, consistency conflicts and communication overhead---contributed to these efficiency differences.

\subsection{Coordination leads to consistency conflicts}

One of the most fundamental challenges that arises from parallel coordination is maintaining \textbf{consistency}: the degree to which different nodes observe a coherent view of shared state despite concurrent updates and communication delays \cite{van2017distributed}.
In order for multiple nodes to make simultaneous progress on a task, distributed systems commonly replicate a shared \textit{state} (e.g., data, files, memory, or intermediate results) across them. However, once multiple replicas of the same information exist, they may be updated independently and at different times. As a result, nodes can temporarily hold different versions of what is intended to be the same underlying information, which must be reconciled to execute the task correctly \cite{lamport1979make, herlihy1990linearizability}.
When one node modifies information that another node is still using, their views of the task can diverge, producing conflicts and downstream errors. 

LLM teams similarly operate over a shared state that must be constantly updated as agents progress, such as task plans, code repositories, analysis outputs, design documents, and evolving context representations. Because multiple agents may read, modify, and use these shared states concurrently, maintaining adequate consistency becomes a central challenge of LLM teams. Otherwise, agents may overwrite each other's work, make changes that break downstream tasks, or make progress with an outdated version of results. 

Without clear protocols for assignment,
we observed three types of consistency violations directly in the decentralized teams in Experiment 2
(Figure~\ref{fig:amdahllaw_decentralized}B). Decentralized teams exhibited a substantial number of concurrent writes, in which two or more agents edit the same file simultaneously and silently overwrite each other's work. We also observed a significant number of rewrites, in which an agent would completely overwrite a file written by a teammate in a previous round. Finally, we observed temporal consistency violations, in which an agent would attempt to implement a task out of order without its predecessor being implemented yet. Because of these consistency conflicts, decentralized teams produced substantially more failed tests (Figure~\ref{fig:amdahllaw_decentralized}C). Across all runs, decentralized teams generated significantly more test failures than preassigned teams (Mann-Whitney $U = 287013$, $p<0.001$). Median failures were 19 for decentralized teams compared to 4 for preassigned teams, and this pattern held within each model individually (all $p<0.001$). These findings are consistent with the distributed systems prediction that decentralized teams will pose a greater risk of consistency conflicts. 

\subsection{Larger and decentralized teams incur compounding overhead}

In order for interconnected nodes to coordinate effectively, they must send packets of information (or \textit{messages}) back and forth to update each other. Sending these messages is not instantaneous, thus incurring substantial costs and overhead. In distributed systems, these communication costs typically depend on how nodes are structured \cite{lynch1996distributed, van2017distributed}. Decentralized architectures require agents to communicate directly with each other, increasing coordination overhead. In the worst case, $n$ agents can induce $O(n^2)$ communication channels, resulting in both delays in processing and a polynomial increase in messages exchanged. In contrast, centralized architectures mediate interactions through a single coordinator, reducing coordination complexity to $O(n)$ as each agent communicates with the central leader. 

\begin{figure}[t]
    \centering
    \includegraphics[width=\linewidth]{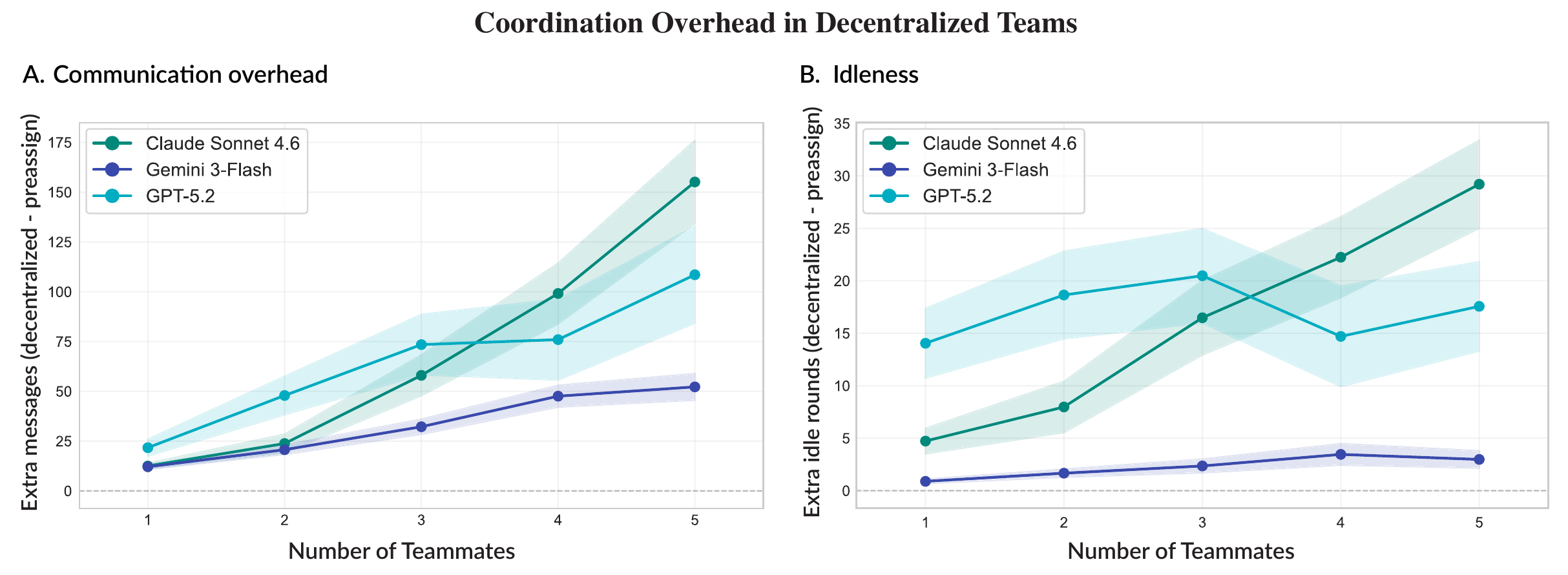}
    \caption{\textbf{Coordination overhead.} Decentralized teams introduce greater coordination overhead, which worsens with more collaborators. \textbf{A. Communication costs:} Each line represents the difference in the number of messages sent when tasks are preassigned vs. decentralized. \textbf{B. Idle costs:} Each line represents the difference in agents remaining idle when tasks were preassigned versus decentralized. Importantly, these agents were still using tokens and sending messages; they just did not complete a task within an idle round. }
    \label{fig:coord_costs}
\end{figure}

Similarly, LLM teams must exchange messages with each other as tasks progress, which should thus theoretically worsen in decentralized teams. To test this correspondence directly, we measured coordination overhead in two complementary ways (Figure~\ref{fig:coord_costs}). First, decentralized teams exchanged substantially more messages than ones with preassigned tasks (Mann-Whitney $U=311551$, $p<0.001$), and these message counts increased with team size ($r=0.483$, $p<0.001$). Second, self-coordinating teams suffered from more \textit{idle rounds}, or interaction steps in which agents communicated but did not complete any task progress (Mann-Whitney $U=289672$, $p<0.001$). These idle rounds reflected delays and contention introduced by peer-to-peer coordination, such as agents trying to work on the same tasks but being blocked or sending messages back and forth about what to do next. Thus, as predicted by prior findings in distributed computing, decentralized LLM teams accumulate substantially more communication and coordination overhead than preassigned teams, which often worsen with more agents. 

\subsection{Decentralized teams mitigate straggler delays}
\label{stragglers}

Another well-documented source of inefficiency in distributed systems is the presence of \textbf{stragglers}: processors that take an unusually long or unpredictable amount of time to complete their assigned task \cite{dean2008mapreduce, zaharia2008improving}. Because many distributed computations synchronize after set phases, overall progress is frequently determined by the slowest workers. In LLM teams, agents must similarly synchronize their outputs (e.g., implementing subtasks before testing, merging results), so downstream progress depends on all agents completing their assigned work.
Variability in reasoning time, tool latency, or context complexity can therefore produce LLM stragglers whose delays stall the entire team. 

\begin{figure}[t]
    \centering
    \includegraphics[width=0.75\linewidth]{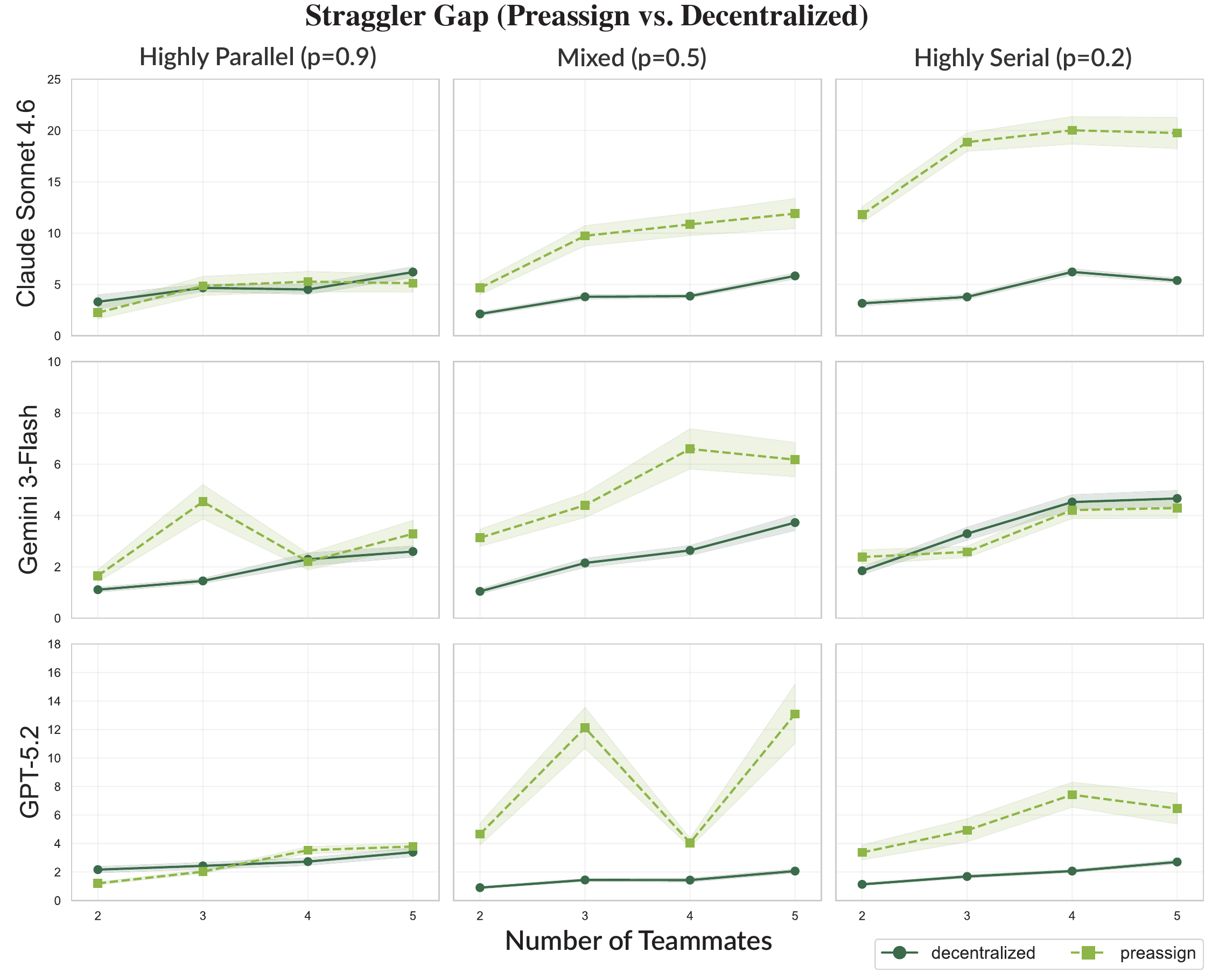}
    \caption{\textbf{Straggler analysis.} When task assignments are fixed (\textit{preassign}), performance is more susceptible to agent variability in the form of \textit{stragglers}: agents that take substantially longer to complete their assigned tasks. This gap arises more often with models that exhibit greater variance in API latency, such as Claude Sonnet 4.6 and GPT-4.1 (see vertical axes), and worsens on mixed or serial tasks where workloads are naturally uneven. When task assignments are decentralized, work can be dynamically reallocated when one agent stalls. The straggler gap is quantified as the difference between the maximum and mean latency within each round, or how many extra seconds the average agent waited for the slowest teammate. Error bars represent standard deviation.}
    \label{fig:stragglers}
\end{figure}

Distributed systems mitigate this problem through replication \cite{wang2015using}. Algorithms like MapReduce duplicate slow or late-stage tasks across multiple workers and accept the earliest completion \cite{dean2008mapreduce, zaharia2008improving}. A similar advantage emerged in self-coordinating teams in Experiment 2 (Figure~\ref{fig:stragglers}). When agents were not locked into fixed assignments, they could flexibly pick up unfinished tasks, allowing faster agents to compensate for slower ones. 
To measure straggler impact directly, we computed a straggler gap, or the difference between the slowest agent’s completion time and the mean completion time of the other agents. This gap was significantly larger in preassigned than decentralized teams (Mann-Whitney $U=8889359$, $p<0.001$), with a median delay of 2.64 seconds compared to 1.42 seconds. The effect was consistent across models (all $p<0.001$), and increased with team size in both coordination modes (preassign: $r=0.23$, $p<0.001$; decentralized: $r=0.35$ $p<0.001$). Finally, this gap was significantly larger for mixed and serial tasks than for parallel ones within the preassign condition (KW $H=137.5$, $p<0.001$; mixed vs. parallel: median $3.91$s vs. $1.73$s, MWU $p<0.001$), consistent with the intuition that a single straggler on the sequential dependency chain forces all downstream work to wait.

This finding provides further evidence that core tradeoffs in distributed systems also emerge in LLM teams. While fixed assignments and centralized coordination can reduce overhead and improve consistency, they are more vulnerable to variability across agents, a disadvantage which becomes particularly pronounced in tasks with greater dependencies. Taken together, these results suggest that LLM teams are vulnerable to coordination challenges that are similar to those faced by distributed systems. In this way, distributed systems offer basic design principles that can be used to understand the tradeoffs faced by centralized and decentralized teams, and to select the team architecture that is best suited for a particular task.

\subsection{Cost-efficiency tradeoffs}

Finally, a long-recognized challenge in distributed systems is that splitting work across machines incurs additional computational costs. Scaling a system requires increased energy consumption and operational resources, resulting in an inevitable tradeoff between performance and energy or budget \cite{gonzalez1996energy}. 

This tradeoff is certainly arising in LLM teams. While certain tasks benefit substantially from LLM teams in terms of wall-clock time, distributing work across multiple agents introduces additional computational costs, including token usage, budget, and energy consumption. Because teams of agents communicate, synchronize, manage consistency, and occasionally cheer each other on (see Appendix 6.2), token consumption often outpaces speedup depending on task structure and model (Table~\ref{tab:scaling}). 

We computed the per-run difference between token multiplier and speedup, where positive values indicate costs outpacing speedup. Preassigned teams showed a small but significant excess overall (Wilcoxon signed-rank $p = 0.013$, median $= 0.02$), driven primarily by the inefficiency of teams performing serial tasks (serial: mean token multiplier $5.83\times$ vs.\ speedup $1.13\times$, $p < 0.001$). 
% For highly parallel tasks, speedup exceeded token costs (median excess $= -0.40$, $p = 0.96$), though this effect depended largely on the model: Claude Sonnet 4.6 achieved a mean speedup of $2.65\times$ at a token cost of only $1.28\times$, while Gemini 3-Flash incurred a $6.98\times$ token increase for a $1.55\times$ speedup. 
Decentralized teams showed a consistently larger gap across all parallelizability conditions (Wilcoxon $p < 0.001$, median $= 1.17$), with token costs scaling strongly with team size (Spearman $\rho = 0.40$, $p < 0.001$) while speedup did not ($\rho = -0.07$, $p = 0.15$). 

These findings highlight that deciding to deploy LLM teams is not purely a performance question in terms of speed or accuracy, but an efficiency-cost optimization problem as well, one that can have huge implications for energy consumption and budgets. Extensions of scaling laws have previously incorporated resource costs when evaluating parallel systems \cite{cassidy2011beyond}. Analogous formulations may therefore provide a useful framework for evaluating LLM team architectures under token or compute budgets. 

\begin{table}[t]
\centering
\caption{Efficiency–cost tradeoffs across task structures and models.}
\label{tab:scaling}
\resizebox{\textwidth}{!}{%
\begin{tabular}{l l cc cc cc cc cc cc cc cc cc cc cc}
\toprule
 & & \multicolumn{6}{c}{Claude-Sonnet-4-6} 
   & \multicolumn{6}{c}{Gemini-3-Flash}
   & \multicolumn{6}{c}{GPT-5.2} \\
\cmidrule(lr){3-8} \cmidrule(lr){9-14} \cmidrule(lr){15-20}
 & & \multicolumn{2}{c}{Parallel} 
   & \multicolumn{2}{c}{Mixed} 
   & \multicolumn{2}{c}{Serial}
   & \multicolumn{2}{c}{Parallel} 
   & \multicolumn{2}{c}{Mixed} 
   & \multicolumn{2}{c}{Serial}
   & \multicolumn{2}{c}{Parallel} 
   & \multicolumn{2}{c}{Mixed} 
   & \multicolumn{2}{c}{Serial} \\
\cmidrule(lr){3-4} \cmidrule(lr){5-6} \cmidrule(lr){7-8}
\cmidrule(lr){9-10} \cmidrule(lr){11-12} \cmidrule(lr){13-14}
\cmidrule(lr){15-16} \cmidrule(lr){17-18} \cmidrule(lr){19-20}
$N$ & & Speedup & Token 
      & Speedup & Token 
      & Speedup & Token
      & Speedup & Token 
      & Speedup & Token 
      & Speedup & Token
      & Speedup & Token 
      & Speedup & Token 
      & Speedup & Token \\
\midrule
\multicolumn{20}{l}{\textit{Preassigned}} \\
\midrule
1 & & 1.00x & 1.00x & 1.00x & 1.00x & 1.00x & 1.00x 
      & 1.00x & 1.00x & 1.00x & 1.00x & 1.00x & 1.00x 
      & 1.00x & 1.00x & 1.00x & 1.00x & 1.00x & 1.00x \\
2 & & 1.80x & 1.57x & 1.50x & 0.97x & 0.76x & 3.96x & 0.59x & 11.76x & 0.60x & 2.73x & 1.18x & 1.63x & 1.12x & 1.84x & 1.26x & 0.74x & 0.96x & 1.49x \\
3 & & 2.42x & 1.06x & 1.14x & 1.46x & 0.59x & 15.39x & 1.13x & 4.98x & 1.26x & 2.51x & 0.68x & 4.25x & 1.60x & 1.26x & 1.10x & 1.89x & 0.61x & 6.24x \\
4 & & 2.78x & 1.16x & 1.48x & 1.36x & 0.65x & 7.02x & 1.64x & 3.63x & 1.47x & 1.93x & 0.53x & 4.60x & 1.44x & 1.89x & 1.25x & 3.03x & 1.02x & 5.66x \\
5 & & 3.35x & 1.33x & 1.30x & 3.04x & 0.79x & 5.76x & 1.14x & 7.53x & 0.94x & 2.72x & 1.12x & 6.87x & 1.36x & 2.48x & 1.14x & 2.55x & 0.96x & 3.40x \\
\midrule
\multicolumn{20}{l}{\textit{Decentralized}} \\
\midrule
1 & & 1.00x & 1.00x & 1.00x & 1.00x & 1.00x & 1.00x 
      & 1.00x & 1.00x & 1.00x & 1.00x & 1.00x & 1.00x 
      & 1.00x & 1.00x & 1.00x & 1.00x & 1.00x & 1.00x \\
2 & & 1.50x & 1.04x & 0.80x & 1.71x & 0.80x & 1.48x & 0.85x & 2.02x & 0.68x & 1.99x & 0.76x & 1.84x & 1.02x & 0.95x & 1.27x & 1.29x & 0.86x & 1.58x \\
3 & & 0.94x & 3.66x & 0.73x & 2.34x & 0.99x & 1.45x & 0.68x & 3.43x & 0.79x & 2.45x & 0.62x & 2.85x & 1.08x & 1.06x & 1.19x & 1.61x & 0.81x & 2.32x \\
4 & & 0.56x & 3.91x & 0.68x & 3.75x & 0.70x & 3.35x & 0.80x & 3.01x & 0.71x & 3.55x & 0.56x & 4.04x & 1.30x & 1.44x & 1.14x & 2.21x & 0.74x & 3.30x \\
5 & & 0.83x & 4.65x & 0.74x & 3.00x & 0.65x & 6.24x & 0.84x & 4.20x & 0.69x & 3.86x & 0.56x & 4.94x & 1.09x & 2.80x & 0.98x & 2.39x & 0.74x & 3.33x \\
\end{tabular}%
}
\end{table}

\section{Discussion}

LLM teams are being increasingly deployed in practice, yet their design remains largely ad-hoc. Teams are assembled, tested, and adjusted through trial and error, with little principled guidance for when coordination will succeed or fail. We argue that distributed systems theory can give us this foundation. The challenges that arise when agents collaborate, such as maintaining consistency, managing synchronization, reducing communication overhead, and absorbing failures, are not novel or unique to LLM teams. They are well-characterized problems that the distributed systems community has spent decades investigating. 

We propose that distributed systems provide a conceptual framework for designing LLM teams, anticipating their limits, and diagnosing how they fail. We demonstrate this framework's utility across two domains. First, we show that the scalability limits predicted by Amdahl's Law arise in LLM teams just as they do in distributed systems (Experiment 1). Second, we show that coordination challenges inherent to distributed systems emerge in LLM teams as well (Experiment 2). Across both, we characterize a fundamental architectural tension between centralized and decentralized team structures, a tradeoff inherited from distributed computing that anyone designing or deploying LLM teams must navigate. Finally, we show that beyond raw speedup, LLM teams introduce meaningful computational overhead in token usage. 

% By viewing LLM teams through this lens, we gain both explanatory insight into observed team dynamics and a generative framework to predict when inefficiencies will arise for a given team design. Such a framework can suggest concrete mechanisms for improving team designs and provide normative benchmarks for evaluating behavior. For instance, this framework can help model builders understand whether there are efficiency gains that can be further eked out by tweaking their design. Scaled up across many teams, such efficiency gains could lead to large cost savings. 

\subsection{Limitations and future directions}

Our experiments provide evidence that supports parallels between LLM teams and distributed systems, but there are many potential avenues for extending this framework in the future. 

Our tasks relied on prespecified dependency structures, which depart from real-world settings where dependencies must be inferred or discovered dynamically. Generalizing to tasks such as text analysis, research synthesis, or open-ended reasoning would test whether the theoretical framework extends beyond structured programming. Future work should also examine complementary scaling laws, including Gustafson's Law, which models performance under scalable workloads, and Gunther's Universal Scalability Law, which captures non-monotonic scaling due to coordination and contention overhead \cite{gustafson1988reevaluating, gunther2008general}.

Additionally, our teams consist entirely of homogeneous agents drawn from the same base model. Prior work suggests that heterogeneous teams, composed of agents with different strengths or base models, can yield accuracy benefits \cite{yang2026understanding}. Formalizing how diversity interacts with coordination and scalability, and exploring if these teams can be optimized with algorithms from heterogeneous load balancing, is a natural extension \cite{ali2005resource}.

Beyond overcoming individual constraints on memory and computation, distributed systems provide an additional benefit: fault tolerance, or the ability to gracefully withstand failures that impact individual nodes \cite{lamport1978implementation, laprie1985dependable}. Systems are designed so that local failures like node crashes, unresponsiveness, or corrupted outputs do not halt overall progress. LLM agents are similarly susceptible to unpredictable faults, including task abandonment, errors, and 
hallucinations that produce arbitrary or misleading outputs \cite{xu2024hallucination, cemri2025multi}. Distributed systems address faults through redundancy, verification, and consensus mechanisms, which should be explored in future work to improve LLM team robustness.

Finally, a related direction concerns the potential application of scheduling and load balancing protocols to improving task assignment \cite{braun2001comparison, selvakumar1994static, sharma2008performance, arpaci2018operating}. As LLM teams are deployed in large-scale settings, they will encounter not only well-defined ``static'' problems but also \textit{dynamic} and partially observable workloads that emerge as tasks are being executed. Ad-hoc negotiation among agents could worsen the coordination overhead and consistency conflicts documented in this paper. Load balancing algorithms from distributed systems offer a principled alternative, potentially improving efficiency by reducing coordination overhead while maintaining utilization across available agents.

\subsection{Conclusion}
Distributed computing provides a rigorous and practical framework for understanding LLM team behavior, generating testable predictions, explaining observed inefficiencies, and pointing to concrete design principles.
The stakes are high in developing such a principled framework for evaluating these multi-agent systems as they are deployed at scale. LLM teams that are poorly coordinated will not merely underperform: they will propagate errors, generate conflicting outputs, and incur substantial costs in terms of compute, energy, and tokens. As these systems scale, these inefficiencies will further compound. We hope that grounding the design of these teams in a formal framework will offer a path towards systems that are not only more capable, but more predictable, efficient, and responsible.

\bibliographystyle{unsrt}  
\bibliography{references}  

\appendix
\section{Appendix}

\subsection{Empirical setup}
\label{empiricalsetup}

\paragraph{Agent architecture}

Each experiment used a team of $N \in \{1, 2, 3, 4, 5\}$ teammate agents, all backed by the same LLM (GPT-5.2, Gemini-3-Flash, or Claude-Sonnet-4-6), with temperature $0.7$ and a maximum of $8{,}192$ output tokens. Each agent maintains a persistent conversation history across all rounds of a run, accumulating task list updates and its own prior replies as context.

Each round, agents receive a system-injected task list showing the current status and owner of all 20 subtasks, then generates a free-text response. The orchestrator parses XML action tags from this response and executes them sequentially. Available actions are \texttt{<claim\_task>}, \texttt{<edit\_file>}, \texttt{<run\_tests>}, and \texttt{<complete\_task>}. All teammate LLM calls within a round are issued concurrently via \texttt{asyncio.gather}.

Agents write code to a shared repository. Concurrent writes are mediated by a file-based pessimistic lock (atomic file creation). If an agent attempts to edit a file already held by another agent, the lock is denied and the failure is returned as feedback. When an agent calls \texttt{<run\_tests />}, the orchestrator runs the full \texttt{pytest} suite and returns the output to that agent as a system message within the same round.

\paragraph{Coordination schemes}

In Experiment 1, agents were preassigned tasks. Tasks distributed programmatically before the run via topological sort and round robin. Each agent receives an explicit system message listing exactly which task IDs are theirs before the run starts, and is instructed not to claim others' tasks. Serial dependency chains were assigned to a single agent. All teammate agents completed their tasks fully in parallel each round. In Experiment 2, agents were asked to self-claim tasks autonomously. 

Every round before acting, each agent receives a system-injected task list showing the current status and ownership of all 20 tasks. Agents cannot claim a task until all its dependencies are marked done, enforced by the orchestrator. Concurrent writes to the shared repo are mediated by a file-based lock (atomic open("x")). A denied lock is returned as feedback to the agent. 

\paragraph{Task set-up}

Each benchmark consisted of twenty subtasks, each requiring the agent to implement a single Python function in its own file. Subtask specifications were fully pre-written, including precise function signatures, docstrings, and expected behavior; agents were instructed not to decompose tasks further. Each subtask file was validated against a pre-written \texttt{pytest} suite that was available to agents via a \texttt{<run\_tests />} action.

We used three thematically distinct benchmarks:

\begin{itemize}
    \item \textbf{MathUtils20}: A numerical utility library. The dependency chain builds a statistical analysis pipeline: 
    \texttt{add} $\to$ \texttt{safe\_add} $\to$ \texttt{sum\_list} 
    $\to$ \texttt{mean} $\to$ \texttt{variance} $\to$ \ldots, where 
    each function imports and delegates to the previous one.

    \item \textbf{DataAnalysis}: A sales data pipeline over a shared hard-coded dataset. The sequential chain builds progressively: 
    \texttt{get\_records} $\to$ \texttt{remove\_invalid} $\to$ 
    \texttt{add\_revenue} $\to$ \texttt{total\_revenue} $\to$ \ldots. 
    Independent tasks implement standalone filtering, grouping, and sorting utilities, each with their own inline sample data.

    \item \textbf{SVGRendering}: An SVG rendering library. The dependency chain begins with low-level formatting primitives 
    (\texttt{fmt\_num} $\to$ \texttt{fmt\_coord}) and builds up to composite elements. Independent tasks implement self-contained shape and element generators (\texttt{make\_rect}, \texttt{make\_circle}, \texttt{make\_line}, etc.).
\end{itemize}

We manipulated task parallelizability by varying the dependency structure among the twenty subtasks. In the \textbf{highly parallel} condition ($p=0.9$, $C=2$), tasks 1--2 form a two-step chain and tasks 3--20 are mutually independent. In the \textbf{mixed} condition ($p=0.5$, $C=10$), tasks 1--10 form a strict sequential chain and 
tasks 11--20 are independent. In the \textbf{highly serial} condition ($p=0.2$, $C=16$), tasks 1--16 form a single sequential chain and tasks 17--20 are independent. The same dependency structure was applied across all three benchmarks. Agents were blocked from claiming a subtask until all its declared dependencies were marked complete by the system.

% \paragraph{Hyperparameters}

% \begin{itemize}
%     \item Temperature: 0.7, max tokens: 8192
%     \item Max rounds: 60 (experiment terminates early if all tasks complete)
%     \item Team sizes: $n \in \{1, 2, 3, 4, 5\}$ teammates
%     \item: Reps per condition: 5
% \end{itemize}

\begin{figure}[h]
    \centering
    \includegraphics[width=0.7\linewidth]{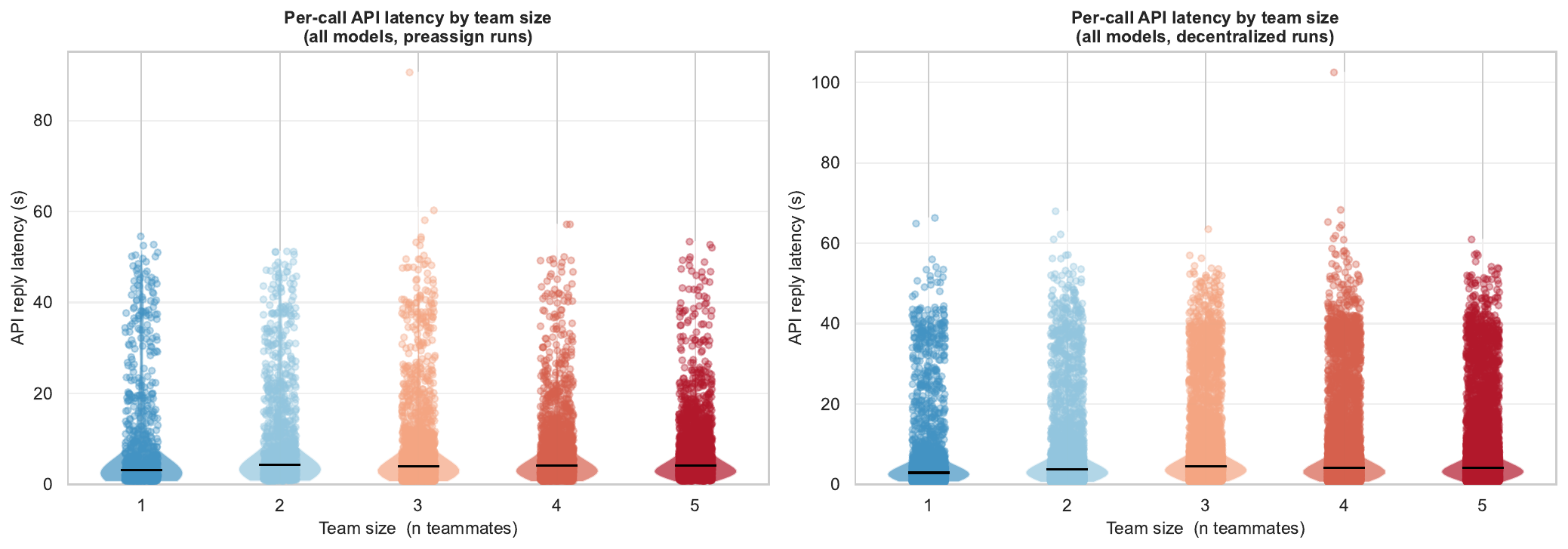}
    \caption{\textbf{API Latency comparison across team sizes.}}
    \label{fig:apilatency}
\end{figure}

\paragraph{Evaluation}

A run was marked successful if all 20 subtasks reached \texttt{done} status within 60 rounds. Task completion was self-reported by agents via \texttt{<complete\_task>}; the system did not independently verify test passage. Each condition ($p \in \{0.2, 0.5, 0.9\}$, $N \in \{1,2,3,4,5\}$) was repeated 5 times. We report success rate (fraction of runs completing all tasks), mean wall-clock time across successful runs, and observed speedup: mean wall-clock time at $N=1$ divided by mean wall-clock time at $N$ teammates. Wall-clock time, rounds to completion, tasks completed, lock conflicts, and total input/output tokens were also recorded per run. Importantly for analysis, there was not large variation in API latency by team size (Figure~\ref{fig:apilatency}).

\paragraph{Exclusion of timed-out runs}

Each experimental trial was subject to a maximum number of 80 orchestration rounds. Runs that did not complete all assigned tasks within this limit were excluded from analysis and repeated up to two additional times; only final failed runs are excluded. Models required the following number of retries due to timeouts (only in the decentralized condition): Claude-Sonnet-4-6: 26, Gemini-3-Flash: 5, GPT-5.2: 60.

\subsection{Examples of inefficient runs}

Agents tended to over-communicate when tasks were mixed or serial to prevent themselves from remaining idle, leading to substantial messaging overhead. For example, in this particular mixed-task run with Gemini-3-Flash, 45\% of replies were wasted messages that yielded no productivity gains, such as:

\begin{itemize}
    \item R15: I am currently idle and waiting for tasks. Let me know if you reassign any tasks to me.
    \item R25: I am still idle since there are no unassigned tasks available. I'm ready to help out if you'd like to reassign any of Dev1's pending tasks to me!
    \item R48: Great job Dev1 on completing task-9! I am still idle as the remaining tasks are all assigned to Dev1. Let me know if you decide to unassign any of them so I can jump in!
    \item R52: Great job Dev1 on completing task-10!
    \item R56: I see that task-15 is now complete! Great job, Dev1.
    \item R57: Looks like Dev1 is working on the final task (task-20)! I'll stand by and cheer them on.
\end{itemize}

A similar pattern emerged with a three-agent Gemini team on a mixed task:
\begin{itemize}
\item R5: I've completed all my assigned tasks. I'll continue to stand by while Dev2 and Dev3 work on their remaining tasks. If I should take over any of their pending tasks to speed up the process, please let me know!
\item R17: I see that Dev2 and Dev3 are currently working on the final two tasks. I will remain on standby until they finish and the project is complete. Great work, team!
\end{itemize}
Serial dependency chains introduced a different failure mode: when one agent became a straggler, others would stall and exhibit highly repetitive behavior. In one run with Claude (team size of 4, decentralized), Dev3 sent the following message 13 consecutive rounds in a row:
\begin{quote}
Still blocked on task-5 (Dev3). All code for tasks 6--16 is written and tests pass. Standing by.
\end{quote}

Coordination failures also arose when agents concurrently claimed the same task and implemented it differently. In round one of a Gemini run (team size of 3), all three agents independently claimed the same task:
\begin{itemize}
    \item Dev1: I have claimed task-1: Implement \texttt{fmtnum}. This function will format a number for SVG use, rounding to \textbf{2} decimal places...   \item Dev2: I have claimed task-1: Implement \texttt{fmtnum}. This function will format a number for use in SVG attributes.
    \item Dev3: ``I have claimed task-1: Implement \texttt{fmtnum}. This function will format a number for SVG use, rounding to \textbf{3} decimal places...
\end{itemize}

\end{document}